\documentclass[prl,twocolumn,showpacs,floatfix,amsmath]{revtex4-1}

\usepackage{color}
\usepackage{xcolor}

\usepackage{graphicx}
\def\be{\begin{equation}}
\def\ee{\end{equation}}
\def\beq{\begin{eqnarray}}
\def\eeq{\end{eqnarray}}

\begin{document}
\renewcommand{\familydefault}{\sfdefault}
\renewcommand{\sfdefault}{cmbr}
\title{Confinement-induced altermangetism in RuO$_2$ thin films}
\author{Samy Brahimi$^1{^*}$\quad Dibya Prakash Rai$^{2,3}$ and\quad Samir Lounis$^{3,4}$ \vspace{0.3em}\\
{\normalsize $^1$Laboratoire de Physique et Chimie Quantique, Universit\'e Mouloud Mammeri de Tizi-Ouzou, 15000 Tizi-Ouzou, Algeria} \quad \\ 
{\normalsize $^2$Physical Sciences Research Center (PSRC), Pachhunga University College, Aizawl 796001, India} \quad \\
{\normalsize $^3$Peter Gr\"unberg Institut, Forschungszentrum J\"ulich \& JARA, 52425 J\"ulich, Germany} 
\quad \\
{\normalsize $^4$ Faculty of Physics and CENIDE, University of Duisburg-Essen, D-47053 Duisburg, Germany}\quad \\
{\normalsize E-mail: samy.brahimi@ummto.dz; s.lounis@fz-juelich.de}
}

\begin{abstract}
The magnetic properties of bulk RuO$_2$ remain a subject of active debate, despite its pivotal role in the emergence of altermagnetism. The latter is a novel paradigm in magnetic phases, characterized by the absence of net magnetization due to anti-parallel alignment of magnetic moments, yet displaying finite spin-splitting in the electronic band structure. This unique behavior unlocks opportunities for advanced applications in information technology devices. Recent experimental and theoretical investigations suggest that bulk RuO$_2$, contrary to prior assumptions, is non-magnetic. In this work, we propose the fabrication of RuO$_2$ thin films to robustly stabilize the altermagnetic phase. Unlike their bulk counterparts, thin films experience substantial strain relaxation, leading to a dramatic impact on the  electronic structure that triggers a transition towards an altermagnetic behavior, which mimics the impact of an artificially applied Hubbard-U correction to account for electronic correlations. Our findings promote the use and exploration of thin films for the realization of spintronic devices based on altermagnets. 
\end{abstract}      

\maketitle
\date{\today}

\textbf{Introduction.} 
Altermagnetism is an emerging magnetic phase that bridges the gap between traditional ferromagnetism and antiferromagnetism, offering a new paradigm in magnetically ordered systems~\cite{Smejkal, Libor, Sinova}. Unlike ferromagnets, where the spin polarization is uniform across the material, or antiferromagnets, where spins are oppositely aligned in a way that cancels the net magnetic moment, altermagnets exhibit a unique collinear magnetic order. This order combines elements of symmetry-breaking and spatially varying spin textures, leading to distinct physical properties.

What sets altermagnetism apart is its spin-space anisotropy coupled with real-space symmetry-breaking, resulting in unconventional spin splitting in the electronic band structure. This splitting lacks spin polarization, distinguishing it from the behavior seen in ferromagnets, and arises without the need for spin-orbit coupling (SOC), a hallmark of traditional antiferromagnetic spintronics. Due to the associated anomalous Hall effects, Nernst effects and spin-split-torques~\cite{Allan, Sinova, Smejkal, Libor, Samanta, Naka, Hayami, Mazin, Gonzalez, Motome, Kusunose, Hernandez, Seo, Ma, Hellenes, Lel, Zhou, Bai, Karub, Hai, Dibya}, altermagnets, present a novel platform for exploring spintronic functionalities that leverage their unique electronic and magnetic properties.

\begin{figure*}[htpb]
\centering
\includegraphics[width=2\columnwidth]{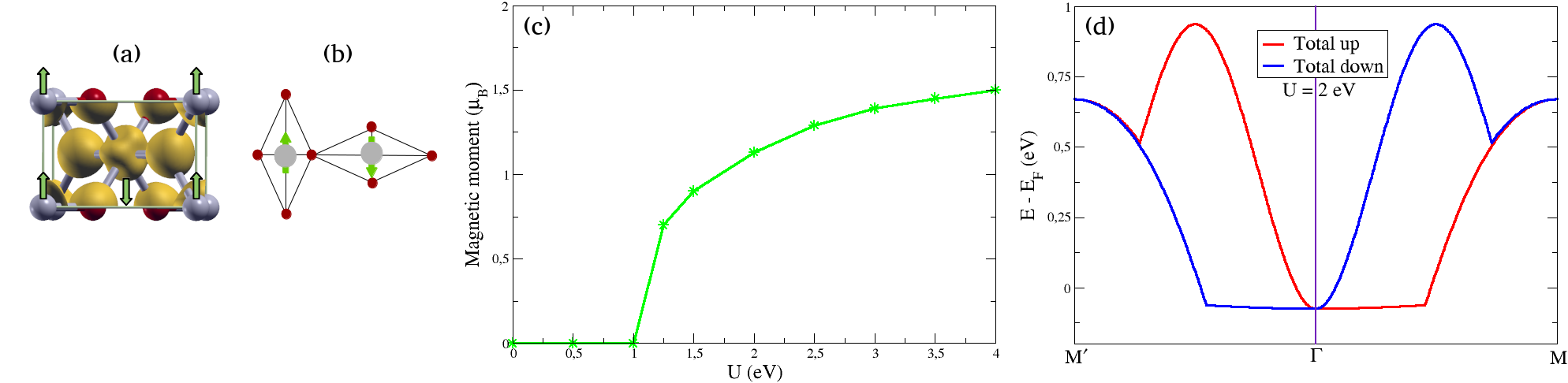}
\caption{Bulk properties of RuO$_2$. (a) The crystal structure of bulk RuO$_2$ showing the antiferromagnetic connection between the two magnetic sublattices and the anisotropic charge density carried by the Ru atoms. (b) Top view of (a) where the Oxygen octaedrons surrounding the Ru atoms are rotated from one to the other based on the spin symmetry $|C_{2}||C_{4}|$ mentioned in the text. Grey and red spheres correspond respectively to the Ru and O atoms.  (c) Magnetic moment of Ru atoms in bulk RuO$_2$ as function of the on-site Coulomb interactions (U). (d) Total NRSS after summation of the bands crossing the Fermi energy $E_F$ as obtained in the bulk phase for the Hubbard U parameter of 2 eV.}
\label{Figure1}
\end{figure*}

RuO$_2$ has been the one of first compounds predicted from first-principles to be altermagnetic (AM)~\cite{Smejkal, Libor, Sinova, Ahn, Hernandez} with magnetic order observed  for temperatures above 300 K~\cite{Berlijn, Zhu}. Altermagnetism in this material finds its origin from the combination between anti-parallel compensated magnetic moments and the roto-reversal symmetry~\cite{Ptok, Allan, Zexin, Gopalan}. As shown in Fig.~\ref{Figure1}(a,b), the two magnetic sublattices in the rutile structure of RuO$_2$ are connected by the so called AM spin-group symmetry $|C_{2}||C_{4}|$, combining a two-fold spin-space rotation with a four-fold crystallographic-space rotation ~\cite{Smejkal}. The rotation of the Oxygen octaedra surrounding the Ru atoms induces  a  charge-density distribution which is not spherically symmetric around the Ru atoms (Fig.~\ref{Figure1}(a,b)), reminiscent of the d-wave symmetry associated to d-wave superconductivity.

Although being extensively studied as a candidate for altermagnetism,   the  magnetic properties of RuO$_2$ remain debated. While early reports suggested Pauli paramagnetism~\cite{Ryden} (confirmed recently~\cite{Wenzel,Kiefer}) or antiferromagnetic order~\cite{Berlijn, Zhu}, first-principles calculations revealed non-relativistic spin splitting (NRSS) up to 1.4 eV under specific Hubbard-U assumptions\cite{Smejkal}. However, realistic (or zero)  U values result in a non-magnetic state~\cite{Smolyanyuk}. It was suggested that magnetic behavior is more likely in non-stoichiometric samples, such as those with Ru vacancies or hole doping while advanced spectroscopy studies~\cite{Philipp, Hiraishi} propose that previously reported magnetic signals likely arose from experimental artifacts rather than intrinsic magnetism. These findings highlight the sensitivity of RuO$_2$'s magnetic properties and the need for precise characterization in future studies.

In this letter, we propose a strategy to design robust altermagnetic (AM) behavior in RuO$_2$ through confinement effects. By reducing the dimensionality of the material and fabricating thin films with varying thicknesses, we demonstrate via ab initio calculations that altermagnetism emerges naturally, without the need for a finite Hubbard-U correction as required in the bulk phase. The driving mechanism is the strong interlayer relaxation in the thin films, which alters the electronic structure, primarily the d$_z^2$ states leading to an energy again,  in a manner analogous to the effect of the Hubbard-U in the bulk. This relaxation induces a magnetic transition characterized by altermagnetic behavior, stemming from the d-wave symmetry of the charge density  already present in the non-magnetic films.

\begin{figure*}[htpb]
\centering
\includegraphics[width=2\columnwidth]{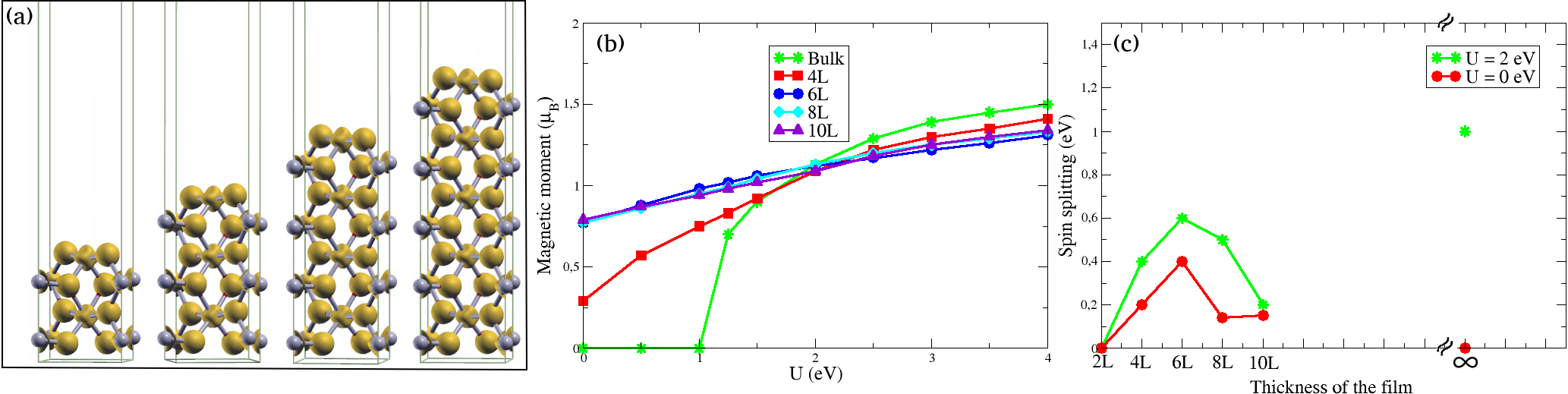}
\caption{(a) Supercells used for the simulation of the (001) RuO$_2$ surface. (b) Average Ru magnetic moments for each film thickness, ranging from 4 layers (4L) till 10L, as a function of the on-site Hubbard U. Total NRSS around the Fermi energy as function of film thickness without and with U (2 eV). The bulk NRSS is also depicted for comparison.}
\label{Figure2}
\end{figure*}

\textbf{Computational details.} 
We carried out first-principles calculations within the framework of DFT+U, assuming the generalized gradient approximation ~\cite{PBE1,PBE2} as implemented in the Vienna ab initio simulation package (VASP), using a plane wave basis and the projector augmented wave (PAW) approach~\cite{PAW1,PAW2}.  
We considered both the of RuO$_2$ bulk phase as well as thin films of different thicknesses along the (001) direction illustrated in Fig.~\ref{Figure2}(a). The energy cut-off for the plane waves is set to 600 eV, while the k-points grids are \(16\times16\times20\)  and \(16\times16\times1\) for respectively the bulk and films. To avoid artificial interactions between periodic images of the films, a sufficient amount of vacuum (15 \AA) is assumed. Atomic relaxations were performed till the forces were less than 0.01 eV/\AA. The optimized bulk tetragonal crystal structure, shown in Fig.~\ref{Figure1}a, is in  agreement with previous works being theoretical or experimental~\cite{Ptok, Mehtougui, Xu, Ze-Jin, Mikami, Zhu}. 
We explore the impact of the Hubbard U and evaluate it via the ansatz of Cococcioni et al~\cite{Cococcioni}. It is introduced in our simulations within the Dudarev's et al~\cite{Dudarev} approach.


\textbf{Results--bulk.} First, we recover the known properties of bulk RuO$_2$ and address the impact of the Hubbard U on the onset of both magnetism and altermagnetism. In Fig.~\ref{Figure1}(c)  we illustrate the evolution of the magnetic moment of Ru atoms as found in the spin-compensated state versus the U-parameter, ranging from 0 to 4 eV,  which includes values assumed   in the  literature~\cite{Sinova, Ptok, Zuntu, Hariki}. 

At low U values, bulk RuO$_2$ remains non-magnetic. A magnetic moment begins to emerge only for U values exceeding 1 eV, eventually saturating at approximately 1.5 $\mu_B$ for U = 4 eV. Using the Cococcioni approach~\cite{Cococcioni}, U is determined to be 2 eV, which clearly triggers a magnetic moment of 1.13$\mu_B$ in agreement with previous ab-initio reports~\cite{Sinova,Ptok}.  
However, the careful analysis conducted recently by Smolyanyuk et al~\cite{Smolyanyuk} on the energetics of differents states obtained for various U values clearly indicates that the U interval [0, 1] eV gives rise to the more favorable phases, which are non-magnetic. This turns out to be in agreement with the experimental report of   Ke{\ss}ler et al.~\cite{Philipp} using muon spin spectroscopy. There is certainly need for more experimental works to understand the origin of the magnetic signal found in earlier experiments~\cite{Tschirner, Olena, Zhu, Berlijn}, where it is suspected that defects and vacancies can trigger a transition towards a magnetic state ~\cite{Smolyanyuk, Diulus, Bolzan, Rogers}. 

The bulk band structure undergoes significant changes as a function of U, an aspect that can, in principle, be readily validated through ARPES experiments. The NRSS is recovered as soon as a moment develops in Ru atoms (for U larger than 1 eV), as shown in Fig.~\ref{Figure1}(d) for the total NRSS after summation of the bands crossing the Fermi energy $E_F$ as obtained for U = 2eV. Clearly, U values larger than 3.5 eV are unrealistic, as they induce a bandgap in the material, which is inconsistent with the well-established metallic nature of RuO$_2$.

\textbf{Results--films.} After examining the bulk phase of RuO$_2$, we turn our attention to (001) thin films of various thicknesses, focusing on those  with an even number of layers (L) ranging from 4L to 10L, to enable the emergence of magnetically compensated phases (Fig.~\ref{Figure2}(a)). The 2L case is excluded from our detailed discussion, as geometric relaxations result in a single monolayer, which does not exhibit AM behavior.

Interlayer distances for each configuration, assuming an initial antiparallel alignment of spin moments, are detailed in Table.~\ref{Table1}. Notably, the films undergo significant structural relaxations: the surface (S) and subsurface (S-1) layers move closer together, while the distance between the subsurface (S-1) and the second subsurface (S-2) layers increases. Furthermore, the reduction in dimensionality substantially impacts the inner-layer distances, underscoring the pronounced effects of confinement on the film's structural  properties.

\begin{table}[h!]
    \caption{Relative change of the interlayer distances in RuO$_{2}$ films with respect to that of the bulk (1.56 \AA). S indicates the surface toplayer.}
    \begin{ruledtabular}
    \begin{tabular}{cccccc}
        Thickness & S, S-1 & S-1, S-2 & S-2, S-3 & S-3, S-4 & S-4, S-5\\ \hline
        2L      & -100\%      &       &      &      & \\
        4L      & -23.7\%      & +24.4\% &   &     &  \\
        6L      &  -19.0\%      & +15,6\%      & -8.8\% & & \\
        8L      & -17.3\%    &  +13.3\%  & -5.1\% & +7.3\% & \\
        10L     & -17.0\%    & +12.7\%   & -3.9\% & +5.3\% & -1.7\%  \\
    \end{tabular}
    \end{ruledtabular}
    \label{Table1}
\end{table}

\begin{figure*}[htpb]                    
\centering
\includegraphics[width=2\columnwidth]{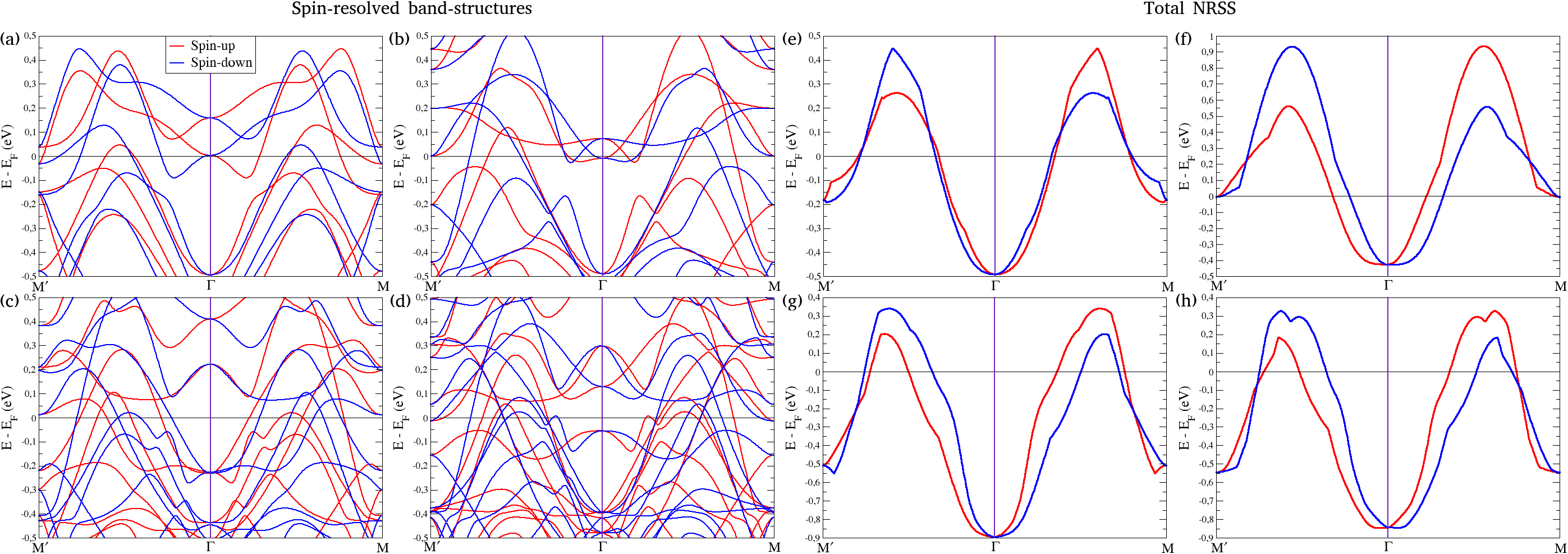}
\caption{Spin-resolved band structures and the corresponding total NRSS (around Fermi energy) for RuO$_2$ thin films as obtained without Hubbard U correction. Results are displayed for films with thicknesses of 4L (a,e), 6L (b,f), 8L (c,g) and 10L (d,h). }
\label{Figure3}
\end{figure*}

In contrast to the bulk phase, all films are magnetic independently from the U values. Since the moments are not necessarily constant across the films, we plot their average value per Ru atom in Fig.~\ref{Figure2}(b). The 4L-thick film carries the weakest moment  (0.25 $\mu_B$) as obtained for zero U. One notices, however, a steady increase of the moments as soon as U is finite, reaching a saturation value of 1.35 $\mu_B$ for U = 4 eV, slightly weaker than that obtained in the bulk phase. We note that for U = 2 eV, the values of the moments of most of the films coincide with those of the bulk. Similar to the latter, the charge density of the films adopts the d-wave symmetry (Fig.~\ref{Figure2}(a)), which ultimately enforces an AM behavior, directly observable in both the spin-resolved band-structure and the summed spin-resolved bands crossing $E_F$ as plotted in Fig.~\ref{Figure3} for U = 0 eV.  
The largest total NRSS  is shown in Fig.~\ref{Figure2}(c) as function of the film thickness without and with U (2 eV). The splitting reaches its maximum for the 6L-thick film, 0.4 eV without U and 0.6 eV with U = 2 eV. 

\begin{figure*}[htpb]                    
\centering
\includegraphics[width=2\columnwidth]{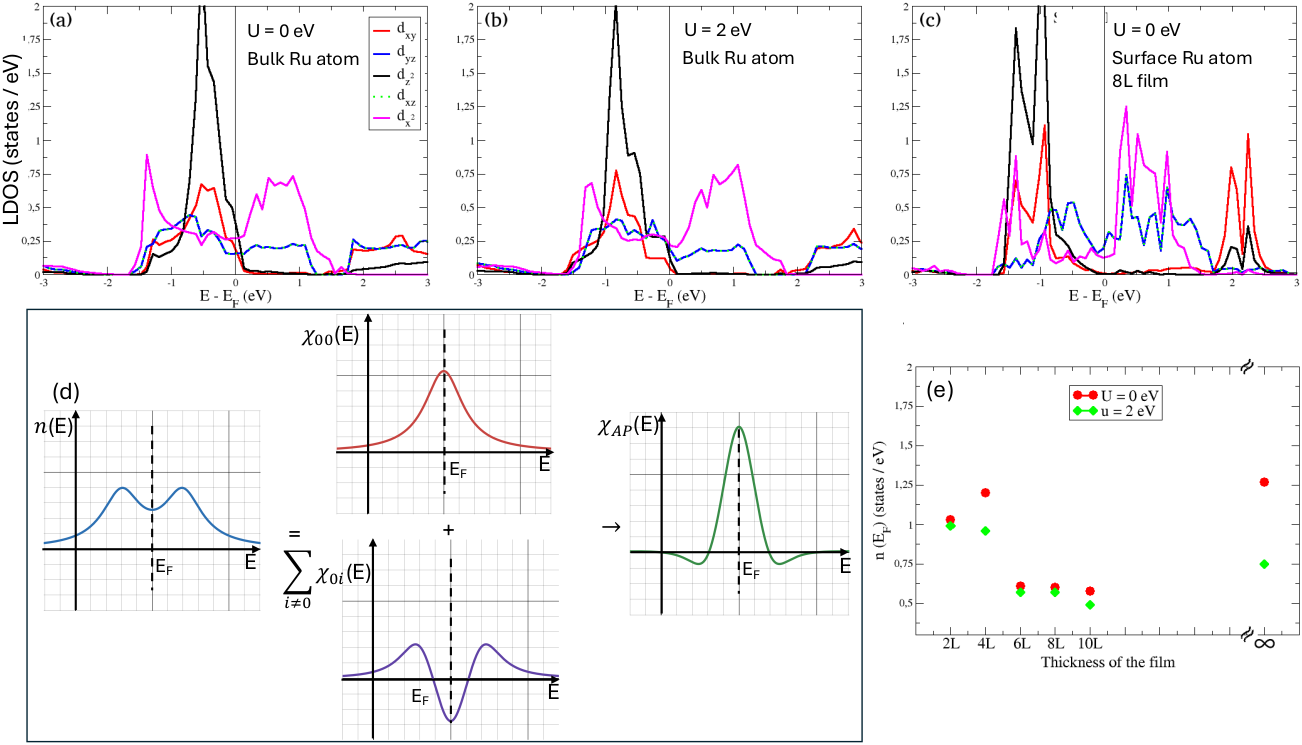}
\caption{Local density of states (LDOS) of RuO$_2$ and approximate criterion for altermagnetism. 
Orbital-resolved Ru LDOS for the different d-states in the non-magnetic phase pertaining to the bulk RuO$_2$ obtained without (a) and with (b) Hubbard U (2 eV), which are compared to that of a Ru surface atom  representative thin film (8L) calculated without U (c). A significant reduction in the LDOS around the Fermi energy is observed in the thin film, resembling the behavior of bulk RuO$_2$ with U, in contrast to the bulk system without U. Notably, the prominent d$z^2$ state (along with d${xy}$) undergoes a substantial energy shift to lower values, driven by strong interlayer relaxations, an effect akin to the application of U. This reduction in LDOS near E$_\text{F}$ enhances the antiferromagnetic susceptibility, facilitating the emergence of the altermagnetic phase. (d) Schematic LDOS of an AFM material and its relation to the intrinsic FM as well as the AP susceptibilities (see text for more details and Ref.~\cite{Bluegel}). The AP susceptibility hosts a large resonance at half-filling, where the LDOS experiences a minimum. (e) the LDOS(E$_\text{F}$) is plotted as function of the films thickness and compared to that of the bulk with and without U. }
\label{Figure4}
\end{figure*}

The analysis of the LDOS of non-magnetic bulk and thin film systems (Fig.~\ref{Figure4}) reveals that the strong interlayer relaxations in the thin films significantly alter the electronic structure of Ru atoms, closely mimicking the effects of the Hubbard U. In particular, the d$_\text{z}^2$ orbital undergoes a pronounced shift to lower energies, driven by the bonding nature of the orbital and the pronounced interlayer relaxations. This behavior is analogous to the impact of Hubbard U in the bulk system. Additionally, this shift is accompanied by a reduction in the LDOS near the Fermi energy, a critical condition for the emergence of a magnetic phase with anti-parallel alignment of spin moments. This phenomenon is further explained below using a simple phenomenological model. The d-wave symmetry of the charge density, dictated by the oxygen octahedral environment, preserves the conditions necessary to induce the NRSS and supports the stabilization of the AM phase. 

\textbf{Phenomenological criterion for anti-parallel spin alignment.}
The Stoner criterion $I \chi_\text{FM} > 1$ describes the condition to form a ferromagnetic (FM) order, which expresses the competition between the intra-atomic exchange interaction $I$ and the kinetic energy in terms of the LDOS $n(\text{E}_\text{F})$ defining the FM susceptibility $ \chi_\text{FM}$ associated to a non-magnetic material. There are materials, however, that when transiting towards a magnetic state prefer an anti-parallel alignment of moments, or even non-collinearity,  instead of being simply ferromagnetic.  To tackle such a case, the Stoner criterion can be generalized, $I \chi > 1$,  to address the instability against the formation of arbitrary magnetic states, each characterized by a specific susceptibility $\chi$~\cite{Gupta,Windsor}. 

A magnetic moment at site $0$ is related to its magnetic neighborhood via 
$\mathbf{M}_0 \approx - I \sum_i  \chi_{0i} M \hat{\mathbf{e}}_i$, where $\hat{\mathbf{e}}$ the unit vector associated to the moment and $\chi_{0i}$  the static  susceptibility connecting sites $0$ and $i$. One can show that $\chi_{0i} = \frac{1}{\pi} \mathrm{Im} \mathrm{Tr} \int^{\mathrm{E}_\mathrm{F}} G_{0i}(E)  G_{0i}(E)$ with $G$ being the Green functions associated to the initially non-magnetic material~\cite{Heine}. Since $\sum_i G_{0i}(E)  G_{0i}(E) = -dG_{00}(E)/dE$, it is straightforward to show that 
$\sum_i \chi_{0i} = n(\mathrm{E}_\mathrm{F})$, which is the response expected for the 
FM response. 

Assuming that the nearest neighbor interaction is the most dominating one, the LDOS are approximately given by~\cite{Bluegel}:
\begin{equation}
    n(\text{E}_\text{F}) \approx \chi_{00} + \sum_{i\neq 0}\chi_{0i}   
\end{equation}
while if the surrounding has moments anti-parallel to moment $0$ the response becomes:
\begin{equation}
    \chi_\text{AP} (\text{E}_\text{F}) \approx \chi_{00} - \sum_{i\neq 0}\chi_{0i} \; ,  
\end{equation}
which is the anti-parallel (AP) magnetic susceptibility.  $\chi_{00}$ is the local atomic susceptibility (obtained before hybridization with neighboring atoms), whose energy dependence is rather simple and follows from atomic Hund's rule-type arguments: The maximum of the moment is expected at half band-filling. 

This is illustrated in Fig.\ref{Figure4}d, where a d-resonance ($\chi_{00}$) of an atom interacting with a bath of electrons undergoes a spectral broadening. Due to the influence of neighboring atoms, orbital hopping occurs, leading to split resonances in $n(\text{E}_\text{F})$, corresponding to bonding and anti-bonding states separated by a minimum. In typical antiferromagnets, the Fermi energy lies at half-filling, where the Stoner criterion for ferromagnetism is less satisfied compared to cases where the Fermi energy aligns with one of the two resonances. The anti-parallel magnetic susceptibility, however, exhibits a large amplitude at the minimum of the LDOS. This framework has been instrumental in explaining the antiferromagnetic behavior of various materials, including Cr multilayers~\cite{Gupta,Windsor,Skriver}. A similar behavior is observed in RuO$_2$, as the LDOS of the non-magnetic compound at the Fermi energy for all films studied without $U$ is weaker than that of the bulk phase (see Fig.\ref{Figure4}e). When $U$ is included, the LDOS diminishes further in all cases, in particular in the bulk, where it reaches a magnitude comparable to that of the films.

\textbf{Conclusions.}
We investigated from ab initio the impact of reduced dimensionality on the altermagnetism of RuO$_2$ considering different thicknesses, going from 2 to 10 layers thick films, which are compared to the bulk phase while accounting for the presence, or not, of the on-site Hubbard U parameter. In contrast to the bulk phase, where the onset of magnetism occurs for values of U larger than 1 eV, the films are found to be not only magnetic independently from U but also  altermagnetic, except for the antiferromagnetic 2-layers thick film.

RuO$_2$ has been the workhorse of altermagnetism but is currently the subject of strong controversies about the origin of its magnetic behavior, if there is any. Our predictions, however, indicate that magnetism, and in particular altermagnetism, is more stable in reduced dimensions, for instance in films, than in the bulk phase. Films lead to confinement effects, which reshuffle the electronic structure by deepening d-states through dramatic relaxations such that magnetism is enabled without the need of the Hubbard U parameter. We utilise a phenomenological model to identify the Stoner criterion for an anti-parallel alignment of the moments ultimately giving rise to the altermagnetic phase.

Our findings promote the exploration and the design of RuO$_2$ films for the emergence of altermangnetism and applications in context of spintronics and information technology.

\newpage
\textbf{Acknowledgments.}
The simulations were performed on the cluster of computers "Aselkam" of the University Mouloud Mammeri in Tizi-Ouzou (UMMTO). S. B. acknowledges funding from the UMMTO and DPR acknowledges financial support from Science \& Engineering Research Board (SERB), New Delhi Govt. of India via File Number: SIR/2022/001150.

\end{document}